\documentstyle{article}

\title{
How to calculate a decoherence matrices numerically
and the microscopic mechanism for decoherence in the Caldeira-Leggett model
}
\author{Hans-J\"urgen Pohle
	\thanks{email: hjp@hpfs1.physik.uni-jena.de}	\\
	\\
	{\it Max-Planck-Gesellschaft}	\\
	{\it Arbeitsgruppe Gravitationstheorie an der Universit\"at Jena} \\
	{\it Max-Wien-Platz 1, D-07743 Jena	}	\\
	\\ \it Germany	\\
}
\begin{document}
\def\theequation{\thesection.\arabic{equation}}
\newcommand{\rf}[1]{(\ref{#1})}
\date{January 1994}
\maketitle
\begin{abstract}
A central object in the interpretation of quantum mechanics of closed systems
with decoherent histories is the decoherence matrix. But only for a
very small number of models one is
able to give explicit expressions for its elements. So numerical methods are
required. Unfortunately the dimensions of this matrices are usually very
high, which makes also a direct numerical calculation impossible. A
solution of
this problem
would be given by a method which only calculates the dominant matrix
elements. This includes to make a decision about the dominance of an element
before
it will be calculated. In this paper I will develop an algorithm that
combines the numerical calculation of the elements of the decoherence matrix
with a
permanent estimation, so that finally the dominant elements will be
calculated only. As an example I apply this procedure to the Caldeira-
Leggett-modell.
\end{abstract}
\pagebreak
\section{Introduction}

The Copenhagen interpretation of quantum mechanics rests on the artificial
division of reality in quantum systems and measuring devices. The result of a
measurement is produced by a collapsing wave function, which is induced by
interaction with a measuring device.  This is
inadequate for a quantum cosmology, where the universe is to be
considered as a quantum system. By definition, there is nothing else
outside the universe dealing as a measuring device. An interpretation of
a quantum theory of closed systems is needed. Such an interpretation has
been developed by Griffiths \cite{Griffiths}, Omnès \cite{Omnes},
Gell-Mann, Hartle \cite{Gell}, Joos, Zeh \cite{Zeh}
and others.  It
makes no reference to external observers, classical apparatus or the
collapse of the wave function.

The main feature of this formulation is to consider a time sequence of
possible outcomes of measuring processes. They are called histories.
Fine-grained or elementary  histories are the basic set of histories,
from which all other coarse-grained histories
can be obtained by unification in the sense of set theory. The way, the
coarse-graining is done depends on the
physical properties of the system one is interested in.

To connect the coarse-grained histories with observables, one has to assign
each of
these coarse-grained histories a probability. The problem is that some of
these histories may have a quantum interference, which violates the sum
rules of the probability theory. These interference's are expressed by means of
a
decoherence matrix. If all the non-diagonal elements of
this matrix are vanishing, the histories will be decoherent. Only in
this case one is able
to assign a probability to each history.

To find the coarse grained sets of decoherent
histories is a central problem of this quantum theory of closed systems.
One can define a decoherence matrix.  Its elements describe the quantum
interference between two histories.

Given the histories, the decoherence matrix can be calculated easily in
principle. The usual type of calculations is simple,
but their number is very large, because the number of possible histories
that is related to the numbers of elements of the decoherence matrix is
usually very
huge. This makes it impossible, to calculate the decoherence
matrix even numerically. For example: Consider a particle in one-
dimensional space divided in $m$ intervals. A coarse-grained history
should correspond to the situation, in which the particle has been
detected in one of these intervals at the times
$\left( t_1,t_2,\ldots ,t_n\right) $.
The number of possible histories would be $m^n$ and
the number of elements of the decoherence matrix would be
$m^{\left( 2n-1\right) }$.

Not all is lost, we are not interested in the decoherence matrix for
any coarse-graining. We are usually interested in such coarse-grainings,
that provide decoherent histories. So if we combine the process of
constructing such coarse-grainings together with the calculation of the
decoherence matrix, we might be more successful. In this case for instance
we already know, that the non-diagonal elements are vanishing.

So we can try to calculate the decoherence matrix in two steps. In the
first step we have to find qualitative features of the decoherence
process to get the orders of magnitude for a coarse-graining that
guaranties the dominance of the diagonal elements of the decoherence
matrix. In a second step we can use these data for a numerical
calculation.

In this paper we want to show how to do these two steps, especially how to
design the numerical calculation for the
Caldeira - Leggett model, which gives us some deeper
insight in the mechanism of decoherence.

\section{
Qualitative features of the Caldeira-Leggett
model
}

This model has been developed by Caldeira and Leggett \cite{Caldeira}, to
explain the Brownian motion. It consists of one distinguished harmonic
oscillator in an interaction with a huge number of other harmonic
oscillators, which we will call the environment. The action of the
distinguish oscillator is
\begin{equation}
   S_A[x]=\frac{M}{2}\int_{0}^{t} dt (\dot x^2 - \omega^2x^2)
\label{eq:wirkung_a}
\end{equation}
and the action of the oscillators of the environment are given by
\begin{equation}
   S_E[R]=\sum_k \frac{m}{2}\int_{0}^{t} dt (\dot R_k^2-\omega_k^2 R_k^2).
\label{eq:wirkung_e}
\end{equation}
The meaning of the variables is obvious. The interaction between the
environment and the distinguished oscillator is given by
\begin{equation}
   S_I[x,R]=-\sum_k \int_0^t dt C_k R_k x.
\label{eq:wirkung_i}
\end{equation}
The $C_k$'s are coupling constants.
At an initial time $t_1$ the density matrix of the whole system should
be
\begin{equation}
   \rho(x_1,R_1,y_1,Q_1;t_1)= \rho_A(x_1,y_1;t_1) \;\rho_E(R_1,Q_1)
\label{eq:rho_0}
\end{equation}
and we assume that the density matrix of the environment is in thermal
equilibrium.
\begin{eqnarray}
   \rho_E(R_0,Q_0)&=&\prod_k
        \exp\bigl[-\frac{m\omega_k}{2\sinh(\omega_k/kT)}
            [(R_k^2+Q_k^2)\cosh(\omega_k/kT)-2R_k Q_k \bigr]
                           \nonumber                     \\
     & &\times \frac{m\omega_k}{2\pi\sinh(\omega_k/kT)}
\label{eq:thermal}
\end{eqnarray}
The space of all possible results of a measuring process is the phase
space $(x,R;p,P)$ with positions $x$ and momenta $p$ of the distinguished
oscillator and $R,P$ are the coordinates and momenta of the environment. An
elementary history corresponds to the classical situation, that at the
times $(t_1,t_2,...,t_n)$ the system can be found at
$((x_1,R_1;p_1,P_1),(x_2,R_2;p_2,P_2),...,
(x_n,R_n;p_n,P_n))$.
We are only interested in the position of the
distinguished oscillator, neither in its momenta nor in the phase space
coordinates of the environment.  In this sense we collect all elementary
histories with equal position coordinates $x_1,x_2,...,x_n$ together in
one family. The most general description
of the above defined families of histories is given by the reduced
density matrix
\begin{equation}
    \rho(x,y;t)=\int dR dQ\delta(R-Q) \rho(x,R,y,Q;t)
\label{eq:reduced_density}
\end{equation}
Now we want to investigate the time development of the reduced density
matrix.  Following the papers by Caldeira and Leggett \cite{Caldeira},and
Halliwell and
Dowker \cite{Dowker} we can represent the evolution of this density matrix with
the help
of a propagator $J$
\begin{equation}
   \rho(x_f,y_f;t) = \int dx_0 dy_0 J(x_f,y_f;t|x_0,y_0;0)\;
   							\rho(x_0,y_0;0) .
\label{eq:propagator_definition}
\end{equation}
By replacing the huge number of oscillators of the environment by a
continuum with the distribution
\begin{equation}
   \rho_D(\omega) C^2(\omega) =
   \left\{ \begin{array}{ll}
      4M m \gamma \omega^2 /\pi &,\omega < \Omega \\
      0                         &,\omega > \Omega
   \end{array}
   \right.
\label{eq:spectral_density}
\end{equation}
(the constant $\gamma$ represents an effective coupling)
the authors could give an explicit formula for the propagator in the high
temperature limit:
\begin{equation}
   J(x_f,y_f;t|x_0,y_0;0) = F^2(t) \exp(iS_{cl} - \Phi)
\label{eq:propagator_explicit}
\end{equation}
where the functions in the exponent are as follows:
\begin{eqnarray}
   S_{cl}=& & K_1\!(t)\; (x_f^2-y_f^2) + K_2\!(t)\; (x_0^2 + y_0^2)
                     \nonumber         \\
          &-&  L(t)\; (x_0 + y_0)(x_f - y_f)
             - N(t)\; (x_f + y_f)(x_0 - y_0)
\label{eq:koeffizienten}
\end{eqnarray}
and
\begin{equation}
\Phi =	A\left( t\right) \left( x_f-y_f\right) ^2	+
	B\left( t\right) \left( x_f-y_f\right) \left( x_0-y_0\right) +
	C\left( t\right) \left( x_0-y_0\right) ^2
\label{eq:Phi}
\end{equation}

The explicit expressions for the time dependent functions
$K_1, K_2, L, N$ and $A,B,C$ will be given in
the appendix, they consist mainly of the exponential function
$\exp(-\gamma t)$
and trigonometric functions in $t$.
This propagator has the remarkable property, that its time dependence
includes only the difference between the initial and final time. Originally one
should have expected, that the propagator is not time-like translational
symmetric, caused by a time dependence of the environment. But by
replacing  the discrete oscillators by a continuum, the  environment became
infinitely large and so the influence of the distinguished oscillator on it
can be neglected.

With the explicit formula \rf{eq:koeffizienten} we have all information about
the evolution
of the density matrix. But the structure of these formulae is very
complex and the high number of parameters makes it difficult to analyze
the evolution in detail. So I decided, to study the decoherence
matrix numerically. The result was, that for a sufficiently
large final time we got asymptotically always the same density matrix.
This reflects the fact, that the oscillator comes to a thermal
equilibrium with the environment.
Now we want to confirm this result analytically.

For the moment we assume, the initial state  of the distinguished
oscillator  at  the time $t_1$ can be described by  the  Gaussian
density matrix
\begin{equation}
   \rho_A(x_1,y_1;t)= V \exp \bigl(
                     - \frac{(x_1-\xi_1)^2}          {\sigma_x}
                     - \frac{(y_1-\eta_1)^2}          {\sigma_y}
                     - \frac{(x_1-\xi_1)(y_1-\eta_1)} {\sigma_{xy}}
                  \bigr).
\label{eq:gaussian_rho_0}
\end{equation}
with arbitrary parameters $ \sigma_x,\sigma_y $ and $\sigma_{xy}$.
This density matrix is normalized by the function $V$. The function
$\rho_A$
represents a Gaussian wave packet in the "squared" configuration space of
the distinguished oscillator. The variables $\xi_0$ and $\eta_0$ denote
complex numbers, which give an information about the position and
"momentum" of this packet. Using the formula \rf{eq:propagator_definition} it
is easy to get the
density matrix at a later time.  One has simply to perform a Gaussian
integration. The result is again a Gaussian concerning the new
position variables $x_f$ and $y_f$.  The coefficients are rational
functions
of the
functions $A,B,C,\ldots ,N$ and thus rational functions of exponential
and trigonometric
functions. So one should assume, that these coefficients can be expanded in a
power series with respect to $\exp(-\gamma t)$ for large $t$.
These calculations are very tedious, but can be carried out by means of
computers. The result is:
\begin{eqnarray}
   \rho_S(x_f,y_f;t)	= V' \exp \Bigl[
                       	&-& \frac{ (\gamma^2 + \omega^2) M}{2kT}
                           		(\frac{x_f + y_f}{2})^2
                     		-\frac{ kTM}{2} ( x_f-y_f )^2
                              					\nonumber      \\
                     	&+&  \exp(-\gamma t ) ( a(t)\; x_f + b(t)\; y_f )
                     			\nonumber		\\
                     	&+& O(\exp(-2\gamma t))
      \Bigr]
\label{eq:expansion}
\end{eqnarray}
The functions $a(t)$ and $b(t)$ describe a harmonic oscillation with the
frequency $\omega$.  The evolution of the density matrix in the squared
configuration space can be splitted in three phases: In the first phase,
which has a characteristic period of $\tau= \gamma t/2$, the shape of the
wave packet relaxes to a certain form, that does not depend on the initial
parameters of the packet at $t_1$. In the squared configuration space this
form represents a two dimensional Gaussian bell which is located over  an
ellipsis $\cal S$. It corresponds to a certain "equi-density" line.
This can be seen from our formula \rf{eq:expansion} by the
fact, that all terms with the coefficient $\exp(-\gamma t)$ become of
smaller order
compared to others. In this sense they can be neglected and thus the
coefficients of the terms quadratic in $x_f$ and $y_f$ become
independent of time and also independent of the initial parameters of
the packet. From formula \rf{eq:expansion}  we can see also,
that the main axes of the
corresponding ellipsis $\cal S$ have a length
of
\begin{equation}
   \tilde d_B^{-2} = \frac{ (\gamma^2 + \omega^2) M}{2kT}
\label{eq:axis1}
\end{equation}
and
\begin{equation}
   d_S^{-2} = \frac{ kTM}{2}.
\label{eq:axis2}
\end{equation}
They are oriented at an  angle of $45$ degrees against the coordinate axes.

While the shape of the packet becomes fixed after the first phase, its
motion, corresponding to a damped oscillation of its position in squared
configuration space is still present.  This can be considered as to be the
second phase. Its characteristic time is also $\tau$. After that time the
system rests on the origin of our coordinate system in the squared
configuration space. This can be considered as the beginning of the third
phase.
In that phase the distinguished oscillator is in thermal equilibrium
with the environment. If we would remove the coupling, e.g. set
$\gamma=0$, then this oscillator will be on equal footing with the
oscillators of the environment. The thermal equilibrium can then be
recognized from our asymptotic formula \rf{eq:expansion} by the fact,
that it is just the high temperature expansion of one of the factors in
\rf{eq:thermal}, but of course with the specific parameters of the
distinguished oscillator. For $\gamma \not= 0$ the whole calculation can
be understood as the quantum analog of the damped harmonic oscillator.

By considering, that all initial functions for the density matrix can be
composed by a linear superposition of Gaussian waves (This system is
even over complete), we see, that the asymptotic behavior, described
above, is very general.

Now we want to use these results, to
construct a physically interesting set of decoherent histories and to
make statements about the corresponding decoherence matrix.
At first we have to enlarge our coarse-graining. From the Heisenberg
uncertainty relation we know that it makes no sense, to measure exact
positions of the distinguished oscillator at the times $t_1,t_2,...,t_n$.
We have to ask questions with a higher degree of coarse-graining. So we
ask only, if the oscillator goes through definite intervals at these
given
times. Our coarse-graining is now, that we collect in one
family all histories,
in which the
position of the distinguished oscillator goes through a sequence of intervals
$\Delta^1_{i_1},\Delta^2_{i_2},...,\Delta^n_{i_n}$ at the times
$t_1,t_2,...,t_n$.
In our notation for the intervals, the upper index means, that we may use
different sets $\Delta^k$ of intervals for different times $t_k$ and the
lower index
denotes which interval of a set we selected. Sometimes we use also the
notation
\begin{equation}
	\Delta_{i_1,i_2...,i_n}=\Delta^1_{i_1},\Delta^2_{i_2},...,\Delta^n_{i_n}.
\label{eq:definition}
\end{equation}
One family is now understood as one of the new coarse-grained histories.
If the sets of intervals for one
particular time are disjunct and their union covers the whole space of
positions, the corresponding
histories will be complete and alternative.

The  decoherence matrix corresponding to the coarse-graining  defined above
is given by
\begin{equation} \begin{array}{l}
\gdef\L1#1{\displaystyle{					#1} \\*[2mm]}
\gdef\Ln#1{\displaystyle{\hspace{1cm} 	#1} \\*[2mm]}
\L1{
	D( \Delta_{i_1,...,i_n}\Delta_{j_1,...,j_n})=
	}
\Ln{
   \int_{\Delta^{n}_{i_{n}} = \Delta^{n}_{j_{n}}} dx_n dy_n \delta(x_n-y_n)
   }
\Ln{
   \int_{\Delta^{n-1}_{i_{n-1}},\Delta^{n-1}_{j_{n-1}}} dx_{n-1} dy_{n-1}
            J(x_{n},y_{n};t_{n}|x_{n-1},y_{n-1};t_{n-1})
   }
\Ln{
   \int_{\Delta^{n-2}_{i_{n-2}},\Delta^{n-2}_{j_{n-2}}} dx_{n-2} dy_{n-2}
            J(x_{n-1},y_{n-1};t_{n-1}|x_{n-2},y_{n-2};t_{n-2})
   }
\Ln{
	\vdots
   }
\Ln{
	\int_{\Delta^{2}_{i_{2}},\Delta^{2}_{j_{2}}} dx_{2} dy_{2}
            J(x_{3},y_{3};t_{3}|x_{2},y_{2};t_{2})
   }
\Ln{
	\int_{\Delta^1_{i_1},\Delta^1_{j_1}} dx_1 dy_1
      J(x_2,y_2;t_2|x_1,y_1;t_1) \rho_A(x_1,y_1;t_1)
   }
\label{eq:decoherence_matrix}
\end{array}	\end{equation}

This  describes a sequence of evolution's between two subsequent moments
of time and following projections. The symbol
$\Delta_{i_1,...,i_n|j_1,...,j_n}$
denotes the set of intervals
$\Delta^1_{i_1},...,\Delta^{n}_{i_n},\Delta^1_{j_1},...,\Delta^{n}_{j_n}$.
We introduce two additional terms: The first is the
incomplete decoherent matrix at time $t_k$. Its elements are defined by the
expression
\begin{equation} \begin{array}{l}
\gdef\L1#1{\displaystyle{		#1} \\*[2mm]}
\gdef\Ln#1{\displaystyle{\hspace{1cm} 	#1} \\*[2mm]}
\L1{\rho_{-\epsilon}(\Delta_{{i_1},...,i_{k-1}|j_1,...,j_{k-1}}|x_k,y_k,t_k) =
	}
\Ln{\int_{\Delta^{k-1}_{i_{k-1}},\Delta^{k-1}_{j_{k-1}}} dx_{k-1} dy_{k-1}
      J(x_{k},y_{k};t_{k}|x_{k-1},y_{k-1};t_{k-1})
   }
\Ln{\int_{\Delta^{k-2}_{i_{k-2}},\Delta^{k-2}_{j_{k-2}}} dx_{k-2} dy_{k-2}
      J(x_{k-1},y_{k-1};t_{k-1}|x_{k-2},y_{k-2};t_{k-2})
	}
\Ln{\vdots
	}
\Ln{\int_{\Delta^{2}_{i_{2}},\Delta^{2}_{j_{2}}} dx_{2} dy_{2}
      J(x_{3},y_{3};t_{3}|x_{2},y_{2};t_{2})
	}
\Ln{\int_{\Delta^1_{i_1},\Delta^1_{j_1}} dx_1 dy_1
      J(x_2,y_2;t_2|x_1,y_1;t_1) \rho_A(x_1,y_1;t_1)
	}
\label{eq:rho_minus_eps}
\end{array}	\end{equation}
A second term is the projected incomplete decoherent matrix at time $t_k$. The
elements are given
by the expression
\begin{equation} \begin{array}{l}
\gdef\L1#1{\displaystyle{		#1} \\*[2mm]}
\gdef\Ln#1{\displaystyle{\hspace{1cm} 	#1} \\*[2mm]}
\L1{\rho_{+\epsilon}(\Delta_{{i_1},...,i_{k}},\Delta_{{j_1},...,j_{k}}
         |x_k,y_k;t_k) = }
\Ln{\qquad      	\left\{ \begin{array}{ll}
         \rho_{-\epsilon}(\Delta_{{i_1},...,i_{k-1}},\Delta_{{j_1},...,j_{k-1}}
               |x_k,y_k,t_k) &,
              x_k \epsilon \Delta^{k}_{i_{k}},
               y_k \epsilon \Delta^{k}_{j_{k}}  					\cr
         0                         &,\mbox{otherwise}
      \end{array}
      \right.}
\label{eq:rho_plus_eps}
\end{array}	\end{equation}
The physically interesting question is, for which sets of intervals are
the corresponding histories decoherent? In all what follow, we assume the
timelike distance between two "measurements" is larger than the
characteristic time $\tau$. The decoherence of the histories is equivalent
to vanishing non-diagonal elements of the decoherence matrix. Any of these
elements corresponds to a pair of histories which have different intervals
at least at one moment of time. Let this time to be $t_k$.
The incomplete decoherence matrix
$\rho_{-\epsilon}(t_k)$ can be written as
\begin{eqnarray}
  & & \rho_{-\epsilon}(\Delta_{{i_1},...,i_{k-1}|,j_1,...,j_{k-1}}
   		                           | x_k,y_k;t_k) =		\cr
  & &\qquad  \int_\infty dx_{k-1} dy_{k-1}
J(x_k,y_k;t_k|x_{k-1},y_{k-1};t_{k-1}) \cr
  & &\qquad \qquad	\times
             \rho_{+\epsilon}(\Delta_{{i_1},...,i_{k-1}|j_1,...,j_{k-1}}
                                    	|x_{k-1},y_{k-1};t_{k-1})
\label{eq:rho_deco}
\end{eqnarray}
This is an evolution of $\rho_{+\epsilon}(t_{k-1})$ from $t_{k-1}$
to $t_k$. If
this time interval is larger than the characteristic time $\tau$, we can
assume that $\rho_{-\epsilon}(t_k)$ is equal to the asymptotic matrix
$\rho_S$.  The subsequent projection of the incomplete decoherence
matrix on the intervals $\Delta^k_{i_k}$ and $\Delta^k_{j_k}$ vanishes
approximately, when these intervals do not catch the ellipses.  This is
guarantied when the distance of the centers of the intervals will be
larger than the width $d_S$ and because these two intervals must be
different and disjunct this is given, when their own width is larger than
that of $d_S$.

If the incomplete decoherence matrix $\rho_{+\epsilon}(t_k)$ vanishes,
then the chain of evolution's and projections in formula
\rf{eq:decoherence_matrix} is cut off and the
corresponding matrix element vanishes also. Because of the criterion, that
the intervals for projections should be larger than $d_S$ does not depend
on any other special property of the histories we considered, it
guaranties decoherence of all histories with that property.  So we will
call $d_S$ the decoherence width from now on.  A remarkable fact is, that
this width does not depend on $\gamma$, measuring the strength of the
coupling. But the timescale for the decoherence does.

In what follows we will always assume, that the widths of the projections
are larger then the decoherence width. So our histories will decohere and
we ask now for their probability. For calculating the probabilities the
projection intervals $\Delta^k_{i}$ and $\Delta^k_{j}$ have to be identical.
We
work in the configuration space of the distinguished oscillator.

Again we start with the incomplete decoherence matrix
$\rho_{-\epsilon}(t_k)$,
which comes from a projection at $t_{k-1}$ and has the asymptotic form of
a wave packet $\rho_S$.  Independent on which interval we project at
$t_{k}$, the asymptotic form, at $t_{k+1}$ will be the same. Only the
prefactors will become different. Are the intervals for projection at
$t_k$ outside the wave packet of $\rho_{-\epsilon}(t_k)$, the prefactor at
$t_{k+1}$ will be zero. Otherwise it will always have approximately the
same order of magnitude if it goes through the packet. This means that we
get approximately the same probability for all histories, whose $k$-th
interval is completely within a region, which is the projection of the main
axis with
$\tilde d^{-2}_B$
of the ellipsis
$\cal S$ to the coordinate axis $x$. The width of that projection is
\begin{equation}
   d_B^{-2}=\frac{ (\gamma^2 + \omega^2) M}{kT}
\label{eq:project_brown}
\end{equation}
If we apply the same argument for all other times, we see that all
histories having intervals only inside a strip of width $d_B$ in the
configuration space of the distinguished oscillator, have the same
probability.
Is the with $d_B$ larger than the asymptotic decoherence width $d_S$, then the
oscillator shows a Brownian motion within an interval of $d_B$.

By analyzing the decoherence width $d_S$ and the width of the Brownian
motion we find the familiar result, that with increasing temperature $T$
the decoherence length becomes smaller, while the Brownian motion
becomes more intense. A stronger coupling makes the Brownian motion
weaker, but has no influence on the decoherence length. A bigger mass
makes the decoherence length and the Brownian motion smaller.

This calculation was for the Caldeira-Leggett model. But the
fact, that in the process of evolution the distinguished
oscillator comes into thermal equilibrium with its environment
in general, makes it very likely that in all similar calculations
we will find a specific asymptotic behavior of the observed subsystem.
This might give the possibility, to set up a quite similar calculation
to get the decoherent histories for other more general systems.

It is also interesting to investigate the evolution of the density
matrix for short times. For our initial density matrix
\rf{eq:gaussian_rho_0} the width for decoherence is:
\begin{equation}
   d_0^{-2}= (\frac{1}{\sigma_x} + \frac{1}{\sigma_y} -
\frac{1}{\sigma_{xy}})/4
\label{eq:d0}
\end{equation}
An expansion of the density matrix for small $t$ gives the following result:
\begin{equation} \begin{array}{l}
\gdef\L1#1{\displaystyle{					#1} \\*[2mm]}
\gdef\Ln#1{\displaystyle{\hspace{1cm} 	#1} \\*[3mm]}
\L1{\rho_S(x_f,y_f;t)	= 	V' \exp \Bigl[
	}
\Ln{- \;(x_f - y_f)^2 \Bigl(
             \frac{1}{d_0^2} + (\  2kTM - \frac{4}{d_0^2 }\  )\ \gamma t \ +
   }
\Ln{\hspace{4cm}  i \frac{( \sigma_x - \sigma_y )
                                             (     \sigma_x \sigma_{xy}
                                                +  \sigma_y \sigma_{xy}
                                                -  \sigma_{x} \sigma_{y} )
                     }
                     { 2M \sigma_x^2 \sigma_y^2 \sigma_{xy} } t \
      \Bigr)
	}
\Ln{- \;\bigl( \frac{x_f + y_f}{2} \bigr)^2
			\Bigl(
             \frac{(     \sigma_x \sigma_{xy}
               +  \sigma_y \sigma_{xy}
               +  \sigma_x \sigma_{y} )
            (M \sigma_x \sigma_y + 2i(\sigma_x - \sigma_y)t)
            }
            {M \sigma_x^2 \sigma_y^2 \sigma_{xy} }
      \Bigr)
	}
\Ln{+ \;  C_x x_f + C_y y_f+ C_0                        \Bigr]
	}

\label{eq:expansion0}
\end{array}	\end{equation}
The variables $C_x$, $C_y$ and $C_0$ represent some time dependent
function in which we are not further interested. We focus our attention on
the real part of the coefficient of $(x_f-y_f)^2$. It describes the
evolution of the
decoherence width $D$. If we replace the temperature by means of the
asymptotic decoherence width $d_S$, see \rf{eq:axis2}, we get
\begin{equation}
   \frac{1}{D^2} =      \frac{1}{d_0^2}
                  + 4 ( \frac{1}{d_S^2} - \frac{1}{d_0^2} ) \ \gamma t
\label{eq:D}
\end{equation}
We see, is the initial decoherence width smaller or larger then the
asymptotic one, the width will be immediately increased or decreased
respectively. So we can say that the relaxation starts immediately.
Suppose the initial width $d_0$ is much larger then the
asymptotic width $d_s$. How long does it take, till the decoherence
width of the oscillator is smaller than a given value of $d$. By assuming
that this width is also much larger than the asymptotic value, we get from
\rf{eq:D}
\begin{equation}
   t_D =  \frac{1}{4\gamma} \bigl( \frac{d_s}{d} \bigr)^2
\label{eq:decotime}
\end{equation}
Depending on the ratio $d_s/d$ this time might be much shorter, than the
relaxation time $\tau$. From formula \rf{eq:expansion0} we see, that the
time dependent part of the width of the Brownian motion is imaginary. So
the range of the Brownian motion is not influenced by the time dependence
in first order. This provides another scenario than described above.  If we
consider coarse-grainings with intervals much larger than the asymptotic
decoherence width, then in a first phase of the scenario the desired
decoherence happens in a very short time compared to the relaxation time.
After that the damped oscillation and Brownian motion, known from the
asymptotic scenario, takes place.

If we choose a coarse-graining with intervals in the order of magnitude
of $d_S$ and the time between measurements in the order of $\tau$, the
diagonal and "near-to-diagonal" elements of the decoherence matrix should be
dominant. So we can neglect the others and the number of elements we
have to calculate is numerically manageable.

\section{How to calculate the decoherence matrix numerically}

In the case of Hamiltonian quantum mechanics the decoherence matrix can
be calculated by a sequence of integration's \rf{eq:decoherence_matrix}.
These integration's
might be carried out numerically. But the number of matrix
elements is very large in practical cases. So, to
get reliable numerical results one has to find out the dominant matrix
elements, without spending to much time for non dominant terms.
Therefore we want to elaborate some inequalities, to give upper bounds
to the matrix elements.

{}From \rf{eq:rho_deco} we get
\begin{eqnarray}
\lefteqn{   D( \Delta_{i_1,...,i_{n}|j_1,...,j_{n}})=
	\int_{\Delta^{n-1}_{i_{n-1}},\Delta^{n-1}_{j_{n-1}}}
		\hspace{-5mm} dx_{n-1} dy_{n-1}
          }                        \nonumber \\
   && \Big[ \int_{\Delta^n_{i_n}} dx_n
      J(x_n,x_n;t_n|x_{n-1},y_{n-1};t_{n-1}) \times \nonumber \\
   &&\qquad
   \rho_{-\epsilon}( \Delta_{i_1,...,i_{n-1}|j_1,...,j_{n-1}}|
   x_{n-1},y_{n-1};t_{n-1}) \ \Big]
\label{eq:numerik1}
\end{eqnarray}
By Cauchy's inequality with respect to the integration's over $dx_{n-1}
dy_{n-1}$
and with the following definitions:
\begin{eqnarray}
\lefteqn{
A(\Delta^{n}_{i_{n}}|\Delta^{n-1}_{i_{n-1}},\Delta^{n-1}_{j_{n-1}})^2=}
			\nonumber		\\
  &&\int_{\Delta^{n-1}_{i_{n-1}},\Delta^{n-1}_{j_{n-1}}}
   dx_{n-1} dy_{n-1}
    \left|  \int_{\Delta^n_{i_n}} dx_n
      J(x_n,x_n;t_n|x_{n-1},y_{n-1};t_{n-1})\right|^2  \nonumber \\
  && \hfill
\label{eq:A}
\end{eqnarray}
and
\begin{equation}
   P^2_{i_{k},..,i_1|j_{k},..,j_1} =
   	\int_{\Delta^{k}_{i_{k}},\Delta^{k}_{j_{k}}}
      				\hspace{-5mm} dx_{k} dy_{k}
         \left|
         \rho_{-
\epsilon}(\Delta_{i_1,..,i_{k}|j_1,...,j_{k}}|x_{k},y_{k};t_{k})
         \right|^2
\label{eq:P1}
\end{equation}
we obtain from \rf{eq:numerik1} the inequality
\begin{eqnarray}
| D( \Delta_{i_1,...,i_{n}|j_1,...,j_{n}}) |  &\leq&
 A(\Delta^{n}_{i_{n}}|\Delta^{n-1}_{i_{n-1}},\Delta^{n-1}_{j_{n-1}})
 \, P_{i_{n-1},..,i_1|j_{n-1},..,j_1}
\label{eq:ineqality1}
\end{eqnarray}
Now, we want to find an upper bound for
$P_{i_{k},..,i_1|j_{k},..,j_1}$. With the help of formula
\rf{eq:rho_deco} we can rewrite equation
\rf{eq:P1} and after rearranging the integrals we get
\begin{equation} \begin{array}{l}
\gdef\L1#1{\displaystyle{					#1} \\*[2mm]}
\gdef\Ln#1{\displaystyle{\hspace{5mm} 	#1} \\*[2mm]}
\L1{P^4_{i_k,..,i_1|j_k,..,j_1} \ \leq
   }
\Ln{P^4_{i_{k-1},..,i_1|j_{k-1},..,j_1}	\times
	\int_{\Delta_{i_{k-1}|j_{k-1} }}
		\hspace{-1cm} dx_{k-1} dy_{k-1}
   \int_{\Delta_{i_{k-1}|j_{k-1} }}
   	\hspace{-1cm} dx'_{k-1} dy'_{k-1}
   }
\Ln{\left| \int dx_k dy_k J(x_k,y_k;t_k|x_{k-1},y_{k-1};t_{k-1})
                             J(x_k,y_k;t_k|x'_{k-1},y'_{k-1};t_{k-1})^*
\right|^2
	}
\label{eq:P4}
\end{array}	\end{equation}
Again by using Cauchy's inequality and with the definition
\begin{equation} \begin{array}{l}
\gdef\L1#1{\displaystyle{					#1} \\*[2mm]}
\gdef\Ln#1{\displaystyle{\hspace{3mm} 	#1} \\*[2mm]}
\L1{  B^4(\Delta^{k}_{i_{k}},\Delta^{k}_{j_{k}}
				|\Delta^{k-1}_{i_{k-1}},\Delta^{k-1}_{j_{k-1}})
        = \int_{\Delta_{i_{k-1}|j_{k-1} }}
        		\hspace{-1cm} dx_{k-1} dy_{k-1}
          \int_{\Delta_{i_{k-1}|j_{k-1} }}
          	\hspace{-1cm} dx'_{k-1} dy'_{k-1}
	}
\Ln{\left| \int_{\Delta_{i_{k}|j_{k} }} dx_k dy_k
         		J(x_k,y_k;t_k|x_{k-1},y_{k-1};t_{k-1})
               J(x_k,y_k;t_k|x'_{k-1},y'_{k-1};t_{k-1})^*
    \right|^2
   }
\label{eq:B}
\end{array}	\end{equation}
we can write
\begin{equation}
   P_{i_k,..,i_1|j_k,..,j_1} \ \leq \ B
(\Delta^{k}_{i_{k}},\Delta^{k}_{j_{k}}|\Delta^{k-1}_{i_{k-1}},\Delta^{k-1}_{j_{k-1}})
   		\, P_{i_{k-1},..,i_1|j_{k-1},..,j_1}.
\label{eq:P5}
\end{equation}
If we know the incomplete density matrix at a time slice $t_k$, we can
calculate $P_{i_k,..,i_1|j_k,..,j_1}$ and an
iterated use of formula \rf{eq:P5} together with \rf{eq:ineqality1} gives
the possibility to give an upper bound for the elements of the
decoherence matrix at any later time. This is not yet very useful for a
numerical calculation, because we can not test the upper bound for all
possible path, caused by their large number. So we have to ask: suppose
we know an incomplete density matrix at a time $t_k$, what is the
upper bound for all decoherence matrix elements independent of subsequent
projections? To answer this question we introduce the quantities
\begin{equation}
\setlength{\arraycolsep}{1mm}
\renewcommand{\arraystretch}{2}
\begin{array}{lcll}
	 A_{n,n-1}(\Delta^{n-1}_{i_{n-1}},\Delta^{n-1}_{j_{n-1}} ) & =
		& 	\max_{\Delta^{n}_{i_{n}}}
		&	\Big[
	  		A(\Delta^{n}_{i_{n}}|\Delta^{n-1}_{i_{n-1}},\Delta^{n-1}_{j_{n-1}})
			\Big]							\\
	 A_{n,n-1} & =
		&	\max_{\Delta^{n}_{i_{n}}|\Delta^{n-1}_{i_{n-1}},\Delta^{n-1}_{j_{n-1}} }
		&	\Big[
	  		A(\Delta^{n}_{i_{n}}|\Delta^{n-1}_{i_{n-1}},\Delta^{n-1}_{j_{n-1}})
			\Big]							\\
	B_{k,k-1}(\Delta^{k-1}_{i_{k-1}},\Delta^{k-1}_{j_{k-1}}) &=
		&	\max _{  \Delta^{k}_{i_{k}},\Delta^{k}_{j_{k}} }\,
		&	\Big[
			B(\Delta^{k}_{i_{k}},\Delta^{k}_{j_{k}}|
			\Delta^{k-1}_{i_{k-1}},\Delta^{k-1}_{j_{k-1}})
			\Big]							\\
	B_{k,k-1} &=
		&	\max_{\Delta^{k}_{i_{k}},\Delta^{k}_{j_{k}},
						\Delta^{k-1}_{i_{k-1}},\Delta^{k-1}_{j_{k-1}}}
		&\Big[
			B(\Delta^{k}_{i_{k}},\Delta^{k}_{j_{k}}|
			\Delta^{k-1}_{i_{k-1}},\Delta^{k-1}_{j_{k-1}})
			\Big]
\end{array}
\renewcommand{\arraystretch}{1}
\label{eq:maxima }
\end{equation}
With these maxima we construct the quantity
\begin{equation}
	K^{(k)}_{i_{k},..,i_1|j_{k},..,j_1}= A_{n,n-1} B_{n-1,n-2}
		,...,B_{k+1,k}(\Delta^{k}_{i_{k}},\Delta^{k}_{j_{k}})
		 P_{i_{k},..,i_1|j_{k},..,j_1} \quad.
\label{eq:K}
\end{equation}
The main point is, that the elements of the decoherence matrix
satisfy the inequality
\begin{equation}
D( \Delta_{i_1,...,i_{n}|j_1,...,j_{n}}) \leq
K^{(k)}_{i_{k},..,i_1|j_{k},..,j_1}
\quad .
\label{eq:DK}
\end{equation}
It says that all matrix elements containing the intervals
$\Delta_{i_{k},..,i_1|j_{k},..,j_1}$ are smaller than the corresponding
value of  $K^{(k)}$, independent of subsequent projections.
Now we can use a computer in the following way: We start with the
initial density matrix $\rho_A(x_1,y_1)$ and calculate the quantities
\begin{equation}
	P^2_{i_1|j_1} = \int_{\Delta^1_{i_1},\Delta^1_{j_1}}
									\hspace{-5mm} dx_1 dy_2 \
									|\rho_A(x_1,y_1)|^2
\label{eq:propgramm1}
\end{equation}
and $K^{(1)}_{i_1|j1}$. We denote with $\cal M$ the set of all
$K^{(1)}_{i_1|j1}$'s and consider the maximum of its elements to be the norm
of $\cal M$
\begin{eqnarray}
	{\cal M}   &=& \cup_{i_1|j_1} \big\{  K^{(1)}_{i_1|j_1} \big\}  \\
	\| {\cal M} \| &=& \max (K^{(1)}_{i_1|j_1})
\label{eq:programm2}
\end{eqnarray}
Now we take the largest element of $\cal M$, the attached intervals
should be $\Delta^1_{I_1} \Delta^1_{J_1}$ and propagate the
corresponding incomplete density matrix. This allows us to calculate the
quantities
\begin{eqnarray}
	P^2_{i_2,I_1|j_2,J_1} = \int_{\Delta^2_{i_2},\Delta^2_{j_2}}
			\hspace{-5mm} dx_2 dy_2
			\ |\rho_{+ \epsilon}(\Delta_{I_1|J_1}|(x_2,y_2;t_2)|^2
\label{eq:programm3}
\end{eqnarray}
and $K^{(2)}_{i_2,I_1|j_2,J_1}$. We enlarge our set $\cal M$ by the
quantities $K^{(2)}$ and reduce it by the element $K^{(1)}_{I_1|J_1}$.
These steps will be repeated: We select the largest
element of $\cal M$, propagate the corresponding incomplete density
matrix and replace this largest element of $\cal M$ by the new obtained
$K$'s. So for instance, we could have after two steps
one of the $K^{(2)}$'s as the largest element or one of the $K^{(1)}$'s.

By repetition of these steps one of our propagation's reach the final timeslice
$t_n$ and we can calculate the corresponding element of the decoherence
matrix. It is not guarantied that this element is the largest, but we
know that all elements we have not yet calculated are smaller than the
norm of $\cal M$. By going on with our procedure we get more and more
matrix elements while the norm of $\cal M$ is decreasing. If this norm
becomes smaller than some desired order of magnitude of the already
calculated matrix elements, we can neglect the not jet calculated
elements and the problem is solved. The only bad thing that can happen
is, that the histories are not decoherent enough. So the program will
produce more and more matrix elements, but the norm of $\cal M$ will
drop down very slowly. If this happens one can find from the calculated
data, which histories are not decoherent and one can combine some of the
intervals to increase the coarse-graining.

To demonstrate the efficiency of the algorithm
we consider the Caldeira-Leggett model. The decoherence
matrix should correspond to the situation where we have $5$ intervals at
$6$ time slices. The parameters of the model and the intervals are
chosen in such a way, that the widths of the intervals are one asymptotic
decoherence width $d_s$ and the distance of the time slices is half of
the relaxation time. With these parameters the decoherence matrix has
$5 \times 5^{5} \times 5^{5} \approx 4.8 \times 10^7$
elements and for each element we have to
calculate $5$ propagation's of the incomplete density matrices. The
time to calculate the decoherence matrix is proportional to the number
of projections
\begin{equation}
P \approx  2.4 \times 10^8.
\label{eq:P}
\end{equation}
Let us assume, that we are not interested in matrix elements
that are smaller than $10^{-1}$ times  the largest element.

Furthermore, for the numerical calculation we arrange the parameters
of the model such,
that the mass of the
oscillator corresponds to that of a proton, the asymptotic decoherence
width should be equal to one atomic length scale $L = 10^{-10} m$ and the
asymptotic width of the Brownian motion will be $3L$.

I made the additional assumption, that the oscillator is in a pure state
at the initial time and that the wave function is a Gaussian function with
a width of $2L$, located at the
origin of the coordinate system.

The program started
to calculate $8478$ elements of the decoherence matrix. Whenever the
program found that the corresponding matrix element will become smaller
than one tenth of the largest element, it terminated that
calculation, so that finally only $1536$ relevant elements had been
determined. The number of propagation's, which had to be performed was
$6025$.
The ratio
between the number of executed propagation's and the total number $P$ is a
measure for the efficiency of the algorithm and has a value of
$2 \times 10^{-5}$.
This numerical calculation, should simply be considered as
a demonstration.

\section{Discussion}
The purpose of this paper has been to provide a guide line for calculating
the elements of the decoherence matrix.
 I  developed an algorithm which
combines the numerical calculation of the elements of the decoherence matrix
with a
permanent estimation, so that finally the dominant elements will be
calculated only.
The algorithm can not be used straight forward to calculate
other models. Because other models require different inequalities to estimate
the matrix
elements. To find these
inequalities may require many analytical calculations - better
estimations provide a more efficient algorithm because
nondominant matrix elements will be ruled out earlier. This has two
advantages: the number of steps to be performed is smaller and
the amount of data to be saved temporarily.
But despite these model-dependent inequalities the main features of the
algorithm remain the same, so that
paper may help to study more complex models.

\begin{appendix}
\section{Appendix}
The explicit formulae for the time-dependent Coefficients in the
exponent of the
propagator $J$ \rf{eq:koeffizienten} and \rf{eq:Phi} are as follows:
\begin{eqnarray*}
a&=&\frac{k T M}{2}\,
	 \csc \left( \omega \tau \right)^2
	 \frac{	 \gamma^2 + \omega^2
	         -\gamma^2 \cos (2 \omega \tau )
	 			-\gamma \omega \sin (2 \omega \tau )
	 			- e^{-2 \gamma\tau } \omega ^2}
			{ \gamma^2 + \omega^2 }
	\nonumber		\\
b&=& k T M \omega \,
   {{\csc (\omega \thinspace \tau )}^2}\thinspace
	\frac{\omega \,\cos (\omega \,\tau )\left( 1-e^{2\,\gamma\,\tau }\right) \,
	      +\gamma\,\sin (\omega \,\tau )
	          \left( 1+e^{2\,\gamma\,\tau }\right) \,}
	     {e^{\gamma\,\tau }\,\left( {\gamma^2}+{\omega ^2}\right) }
	\nonumber		\\
c&=& \frac {k T M}{2}
	\thinspace
	{\csc (\omega \tau )}^2
	\thinspace
	\frac {- \gamma^2 - \omega^2 + e^{2 \gamma \tau}\,{\omega^2}
	       + \gamma^2 \cos (2\,\omega \,\tau )
	       - \gamma \omega \sin (2\,\omega \,\tau )}
		   {\gamma^2 + \omega^2}
	\nonumber		\\
\label{eq:abc}
\end{eqnarray*}
and
\begin{eqnarray*}
K_1&=&-\frac{M\gamma}2+\frac{M\omega }2\cot \left( \omega \tau \right)
	\nonumber		\\
K_2&=&+\frac{M\gamma}2+\frac{M\omega }2\cot \left( \omega \tau \right)
	\nonumber		\\
L&=&\frac{M\omega }{2\sin \left( \omega \tau \right) }e^{-\gamma \tau }
	\nonumber		\\
N&=&\frac{M\omega }{2\sin \left( \omega \tau \right) }e^{\gamma \tau }
	.
	\nonumber		\\
\label{eq:KLN}
\end{eqnarray*}
\end{appendix}
%
\centerline{\bf  Acknowledgments}
\vskip .26 in
I am grateful to the Department of Physics of the
University of California/Santa Barbara, were part of this work was
carried out and the Deutsche Forschungsgemeinschaft for supporting this
research by funds.
I would particularly like to thank Jim Hartle for many useful
conversations over a long period of time.
%

%
\end{document}